# Ubiquitous Indoor Localization and Worldwide Automatic Construction of Floor Plans

Moustafa Youssef, Moustafa Elzantout, Reem Elkhouly, Amal Lotfy
Egypt-Japan University of Science and Technology
Alexandria, Egypt

**Abstract**

Although GPS has been considered a ubiquitous outdoor localization technology, we are still far from a similar technology for indoor environments. While a number of technologies have been proposed for indoor localization, such as WiFi-based techniques, they are isolated efforts that are way from a true ubiquitous localization system. A ubiquitous indoor positioning system is envisioned to be deployed on a large scale worldwide, with minimum overhead, to work with heterogeneous devices, and to allow users to roam seamlessly from indoor to outdoor environments. Such a system will enable a wide set of applications including worldwide seamless direction finding between indoor locations, enhancing first responders' safety by providing anywhere localization and floor plans, and providing a richer environment for location-aware social networking applications.

We describe an architecture for the ubiquitous indoor positioning system (IPS) and the challenges that have to be addressed to materialize it. We then focus on the feasibility of automating the construction of a worldwide indoor floor plan and fingerprint database which, as we believe, is one of the main challenges that limit the existence of a ubiquitous IPS system. Our proof of concept uses a crowd-sourcing approach that leverages the embedded sensors in today's cell phones, such as accelerometers, compasses, and cameras, as a worldwide distributed floor plan generation tool. This includes constructing the floor plans and determining the areas of interest (corridors, offices, meeting rooms, elevators, etc). The cloud computing concepts are also adopted for the processing and storage of the huge amount of data generated and requested by the system's users. Our results show the ability of the system to construct an accurate floor plan and identify the area of interest with more than 90% accuracy. Further algorithmic processing is expected to increase the system's accuracy. We also identify different research directions for addressing the challenges of realizing a true ubiquitous IPS system.

## 1 Introduction

Many indoor location determination technologies have been proposed over the years, including: infrared [18], ultrasonic [16], and radio frequency (RF) [4]. All these technologies provide varying levels of accuracy that can support different application needs. However, each of these technologies is designed to be deployed in a certain area, with known floor plans, and some of them require deployment of special hardware and/or require special calibration of the area of interest.

In this paper, we address the problem of realizing a ubiquitous indoor positioning system (IPS). Similar to the outdoor GPS, a ubiquitous indoor positioning system is envisioned to be deployed on a large scale worldwide, with minimum overhead, to work with heterogenous devices, and to allow users to roam seamlessly from indoor to outdoor environments. Such a system will enable a wide set of applications including worldwide seamless direction finding between indoor locations, enhancing first responders' safety by providing anywhere localization and floor plans[1], and providing a richer environment for location-aware social networking applications.

The IPS system we envision is based on leveraging the cell phone as a ubiquitous computing device with a number of internal sensors, such as accelerometers, compasses, and cameras. Figure 1 gives an overview of an IPS system components. The system consists of five main modules: (1) a signal collection module that collects raw data from the sensors on the cell phones, (2) a floor plan construction module that constructs a worldwide floor plan database, (3) a location information database builder module that collects the information needed for localization, (4) a location estimation engine, and (5) a user interface module that receives users queries and return the floor plans and the estimated users' locations. The cloud computing concepts will be leveraged for the processing and storage of the huge amount of data generated and requested by the system's users.

Among these modules, we believe that obtaining floor plans for virtually all buildings worldwide is the main challenge that limits the existence of a ubiquitous indoor localization system. As an example, although WiFi has been installed in a large number of buildings worldwide and WiFi localization can give meter accuracy [20], WiFi localization cannot be leveraged as there are no floor plans for all WiFi-enabled buildings. This can be due to the lack of building blueprints, as is the case in many developing countries, privacy issues, or to the fact that no one is willing to take the effort to submit the floor plan to a worldwide floor plan database. Even if someone submits the floor plan, there is still a manual effort required to process the floor plan to identify the areas of interest. In addition, a floor plan cannot be used for WiFi localization without constructing a fingerprint for the area of interest, which is a time consuming process. Updating the floor plan and the fingerprint is a task that has to be repeated from time to time to capture changes in the environment. Clearly, a true worldwide ubiquitous localization system has to overcome these hurdles. We believe that automating the floor plan construction process and further the fingerprint generation process is one of the main challenges that need to be addressed first. This has to be combined with other challenges in order to meet the vision of a true ubiquitous localization system. These include: seamless roaming between indoor and outdoor environments, user privacy issues, energy-efficiency aspects, devices' heterogeneity, massive data processing, and incentive models. Combine all of these challenges with the scalability issues, in terms of devices generating data and devices requiring location information, and the challenges get a new dimension.

---

[1] A number of commercial systems for indoor direction finding have started to emerge, e.g. Point Inside and Micello Indoor Maps. Such systems depend on manually building the floor plan.

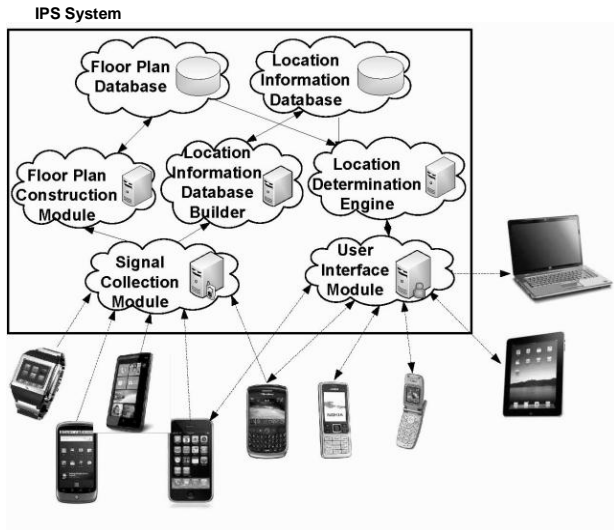

Figure 1: The IPS system architecture.

Therefore, in this context, we make three contributions is this paper:

- We define a vision for a ubiquitous indoor positioning system.
- We study the feasibility of automating the process of constructing a worldwide floor plan database. In particular, we seek to find answers to two main questions: (1) by using the raw sensor data from today's smart phones, can we construct reliable user traces? and (2) based on the constructed user traces, can we identify the areas of interest on the floor plan?
- We identify and discuss several research challenges that

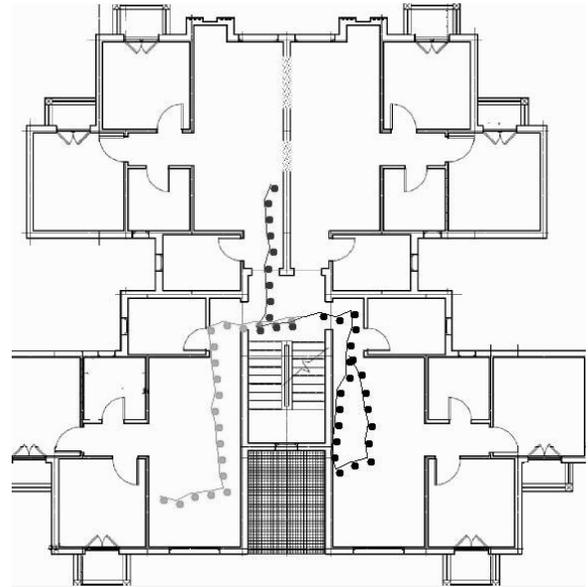

Figure 2: An example of the different traces generated by the movement of different users using the cell phones' inertial sensors.

exist in realizing both the algorithms and infrastructure support.

The rest of the paper is organized as follows: Section 2 introduces the IPS vision and identifies the different sub-components of an IPS system. Section 3 describes a number of algorithms that show the feasibility of automatically constructing a floor plan using cell phone sensors. Section 4 identifies several challenges in realizing the IPS vision. Section 5 discusses related work, and finally Section 6 summarizes the paper.

## 2 Overview and Challenges

In this section, we describe our vision for the IPS system based on Figure 1.

Our vision is based on a crowd-sourcing approach where sensor information are collected from cell phones that can be used to construct user traces inside buildings. Such traces describe areas inside the building that the user can move in and hence can be used to construct a floor plan (Figure 2). The user traces along with the sensor data can be further processed to estimate the different areas of interest inside the building, including, rooms, corridors, elevators, meeting rooms, etc. Fingerprint construction for localization can be constructed concurrently as the data is collected and the floor plan is materialized. We believe that RF technologies, such as WiFi or GSM, would be a good candidate for a ubiquitous localization system as they are available in a large number of buildings worldwide and in a large number of personal devices (such as cell phones, tablets, embedded devices, etc). The cloud computing services are used to scale the system to a large number of users and to a worldwide scale. For the rest of this section, we describe the details of each module assuming a single building for simplicity.

### 2.1 Raw Signal Collection Module

This module is responsible for collecting the data from the cell phones' sensors. These sensors may produce different signals such as the acceleration, compass readings, gyroscope readings, WiFi information (signal strength and heard APs), GSM information (signal strength and nearby cell towers), camera images, sound signals, outdoor GPS signal, etc. Each signal is associated with a timestamp and is forwarded to other modules for processing.

### 2.2 Floor Plan Construction Module

This module takes the raw and processed sensor signals from the large number of users who use a building daily and fuse them to estimate the floor plan. It is also responsible for dynamically updating the floor plan. The estimated floor plans are stored in the cloud for later retrieval. Our current prototype has two submodules: the user trace construction module and the area of interest detection module.

#### 2.2.1 User Trace Construction Module

This module is responsible for the initial processing of the raw data, such as estimating the velocity and displacement from the other sensors. It combines both the displacement with the direction information from the compass to estimate the user trace inside the building. More details about this module can be found in Section 3.2.

#### 2.2.2 Area of Interest Detection Module

Once the user traces are generated, they are fed to this module to detect the different areas of interest and thus construct the floor plan. It uses the raw and processed signals as features for detecting the areas of interest. More details about this module can be found in Section 3.3.

### 2.3 Location Information Database Builder Module

This module is responsible for constructing the fingerprint database for the location determination system. It populates the fingerprint database, also stored in the cloud, based on the WiFi and GSM information it gets from the cell phones the users use inside the building. The challenge here is how to associate the collected data with the user location, when the user location is not known. Different approaches can be used as discussed in Section 3.4. Note that other localization technologies can also used with the system as we discuss in Section 4.10.

### 2.4 Location Determination Engine

This module is responsible for estimating the user location based on the signals it receives in the user query. It consults the fingerprint generated by the Location Information Database Builder Module.

### 2.5 User Interface Module

This module is responsible for interacting with the system users. It receives a query from the users about their current location. This query may include a variety of sensor information, such as WiFi information, camera image, and other signals. The module then consults its location estimation engine and returns the current estimated user location and possibly the floor plan the user is currently located at.

### 2.6 Goal

After defining the different subfunctions of the IPS system, the next contribution of this paper is to study the feasibility of the the floor plan construction module, as we believe it is the main challenge for realizing a ubiquitous IPS vision. In addition, we define other research challenges that need to be addressed for the IPS problem to become a practical system taking into account the other modules.

## 3 Floor Plan Construction

In this section, we provide actual experimental results to demonstrate the feasibility of the floor plan construction along with algorithms for the user trace construction and areas of interest identification. We defer the discussion of the other IPS modules to the next section. We start by describing the environment we used for our testbed.

### 3.1 Test Environment

As an example of a typical environment for the IPS, our experiment is conducted in a floor plan covering an area of $448m^2$ and covered with an 802.11 network. We have three different users carrying three android phones (two Nexus One phones and one Samsung Galaxy S phone). All phones have the same set of sensors: 3-axis accelerometer, 3-axis magnometer, WiFi signal strength and heard APS MAC addresses, GSM signal strength and heard towers IDs, and GPS. Although the GPS does not work indoors, we use it as a synchronization point to determine global reference points. In addition, even though the phones contain cameras and microphones for capturing images and sound signals respectively, these are not used in our current testbed. The camera and mic can be exploited to determine areas of interest as in [3].

Sensors are queried using SENSOR DELAY GAME mode and data is transmitted to a server along with their timestamp and user ID. Each one of the three users generates random traces over different days and covers different areas of the floor plan. Figure 2 shows the layout of the experiment.

### 3.2 User Trace Construction

The first step in constructing the floor plan is to construct the user traces inside a building from the raw sensor data (Figure 2). The idea is to obtain global reference points from the sensors, when possible, and switch to relative localization when such global reference points are not available. Therefore, this task can be further decomposed into: Obtaining global reference points and estimating the user location.

### 3.2.1 Obtaining Global Reference Points

This function refers to determining reference points that can be used to map a local floor plan to a global coordinate system. In addition, they can be used to reduce the error in the constructed traces as discussed below. Obtaining global reference points is achieved in our prototype using different techniques:

1. The last reported GPS location before entering the building, detected by the loss of the GPS signal, is used as a global reference point. Similarly, reference points can also be obtained near windows.

2. APs locations can be used as reference points. These locations can be estimated as the points in the floor plan where the AP signal strength is maximized. Note that the AP can be installed at a different floor. The AP location we refer to here is a virtual relative location within the floor plan just enough to be used as a reference point. The coordinate of the AP can be estimated based on the reference points obtained by method 1 and the relative displacement to these points.

3. Intersection between different user traces, or traces from the same user but at different times, can be used to determine global reference points on the traces of users with no GPS devices.

4. Similarly, the images and surrounding sound signature, when available, can be used to obtain synchronization points between different traces.

5. Some areas of interests, such as stairs and elevators, can be used as global reference points once identified. These areas are easy to identify, as we show in Section 3.3.4, and provide a good way to identify reference points when the user changes floors.

### 3.2.2 Estimating the User Location

Once the user is inside the building, there is no GPS signal. Therefore, other ways for localization need to be used. To be able to perform localization in any indoor environments without requiring installation of special infrastructures, we rely on a dead reckoning based approach. In dead reckoning the current location ($X_k, Y_k$) is estimated with help of the previous location ($X_{k-1}, Y_{k-1}$), distance traveled ($S$) and direction of motion ($\theta$) since the last estimate as:

$$X_k = X_{k-1} + S * \cos(\theta) \quad (1)$$

$$Y_k = Y_{k-1} + S * \sin(\theta) \quad (2)$$

$\theta$ can be estimated from the magnetometer, while the displacement $S$ can be obtained from the accelerometer.

**Displacement Estimation** Theoretically the distance traveled can be calculated by integrating acceleration twice with respect to the time. However due to the presence of noise in the accelerometer output, error accumulates rapidly with the time. Another source of error is the presence of a component of acceleration due to the gravity of the earth. These factors lead to errors in position that will grow cubically with the time and can reach 100 meters after 1 minute of operation [19].

To avoid accumulation of errors, our approach is based on the zero velocity update technique (ZVU) [13]. The ZVU makes use of the fact that the acceleration signal during the human motion will have periodic zero values for velocity. These points of zero velocity can be used for synchronization and thus makes error in displacement linear in time rather than cubical. This is equivalent to using the accelerometer as a pedometer. Therefore, the total distance traveled is estimated as the sum of the individual step sizes. Figure 3 shows the magnitude of acceleration during the process of walking for five steps. The figure shows that each step is represented by a positive peak followed another negative acceleration peek.

To detect the number of steps, we started by implementing the local variance threshold method [10], which was reported to give an amazingly 0.1% error in number of detected steps. The method is based filtering the magnitude of acceleration followed by applying a threshold on the variance of acceleration over an a sliding window. Unfortunately, as Table

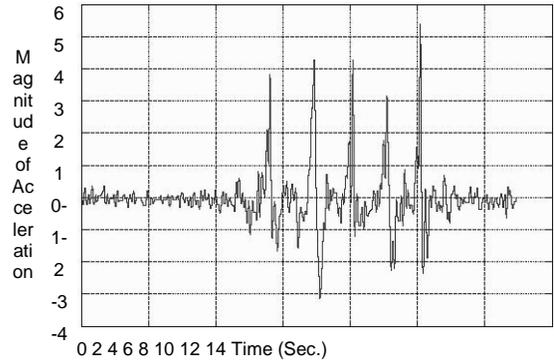

Figure 3: An example of the acceleration pattern for five steps.

| Actual Steps | FSM-based | Local Variance |
|---|---|---|
| 300 | 297 (1.0%) | 45 (85.0%) |
| 270 | 278 (2.9%) | 217 (19.6%) |
| 120 | 121 (0.8%) | 74 (38.3%) |
| 19 | 19 (0.0%) | 12 (36.8%) |
| 11 | 10 (9.0%) | 2 (81.8%) |
| 4 | 4 (0.0%) | 6 (50.0%) |
| Average Error | 2.3% | 51.9% |

Table 1: Performance of the proposed FSM-based method for detecting the number of steps under different traces. Number between parenthesis represent percentage of error.

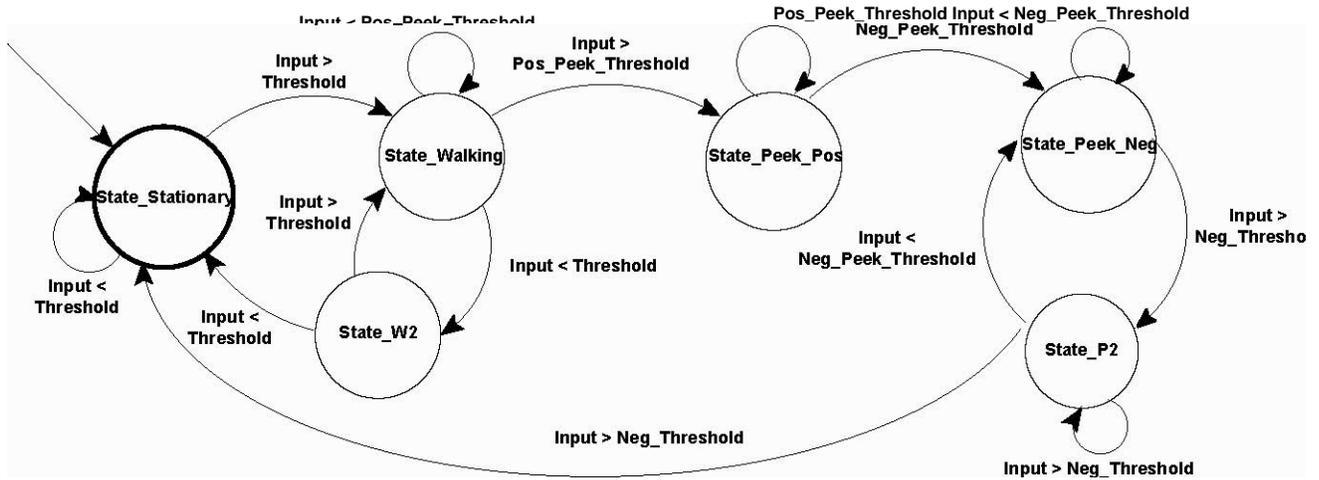

Figure 4: The finite state machine for detecting the number of steps from the accelerometer input.

1 shows, the method performed poorly in our testbed with an average percentage error of 51.9%. The table also shows that this error is independent from the number of steps taken. We believe that this is due to the difference in sensors, as the inertial sensors where attached to the foot in [10] and to the variance and more noise of the motion patterns in our traces as compared to those in [10].

To address the noise in the acceleration signal and to further enhance the accuracy of step count detection, we introduced a **new step detection** method using a finite state machine (Figure 4). there are two main states: (1) State Stationary representing the case when the user is not walking and (2) state State Walking representing the case when the user is currently making a step. Another two accessory states are used: (1) State Peek Pos indicating that a positive peek has been reached and (2) State Peek Neg indicating that a negative peek has been reached. Two more states are used to tolerate the noise and outlier measurements: (1) State P2 and State W2.

In order to detect a step, the following states have to be passed through during the state transition (State Stationary, State Walking, State Peek Pos, State Peek Neg, State Stationary). The step starts when a transition from State Stationary to State Walking happens. The moment of transition again to State Stationary marks the end of the step. **Angle Correction** To be able to estimate the new location after a step has been taken we also need to know the heading, i.e. direction of motion. This can be estimated by using the measurement of the magnetometer. However since the output of the the magnetometer is very noisy, we apply the heuristic that the change of the heading angle between steps can only be a multiple of 45 degrees. This assumption is based on the fact that a person makes only big change of heading when making a turn, such as entering a room or switching direction.

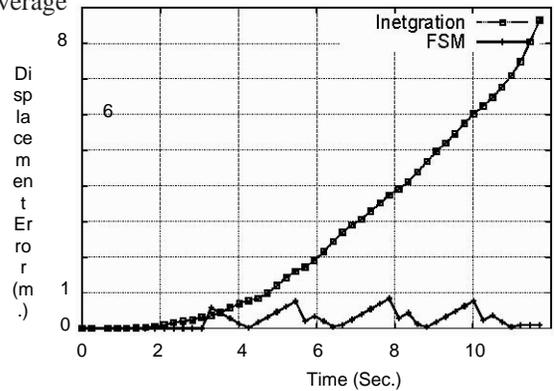

Figure 5: Comparison between displacement error using the proposed FSM approach and the integration approach."

### 3.2.3 Results

Table 1 shows the error in estimating the number of steps when using the proposed FSM method as compared to the variance based method. The table shows that the proposed technique can achieve an average accuracy of 2.3% in estimating the number of steps under different scenarios. Figure 5 shows the displacement error for a 7-step scenario . The figure shows that or technique leads to a maximum displacement error of 0.9m. On the other hand, using the standard integration technique leads to a significant error in displacement during the same period due to the accumulation of error.

### 3.3 Areas of Interest Estimation

Once the user traces are constructed, the next step is to construct the floor plan and identify the areas of interests. Our technique is based on dividing the unknown floor plan area into blocks and classifying each block based on the traces that go through it (Figure 6). We have four different classes: Office, corridor, elevator, and stairs. Note that the building outline can be obtained though different techniques including:

- As the outline of all traces collected by all users using the floor plans of the building. This can be expanded dynamically as more traces become available.
- By applying the same idea of using user GPS traces in outdoor environments to detect the building outline.
- By using a mapping service, such as Google and Bing Maps. GPS coordinates outside buildings along with the satellite imagery and/or the road map can be used to detect the building outline.

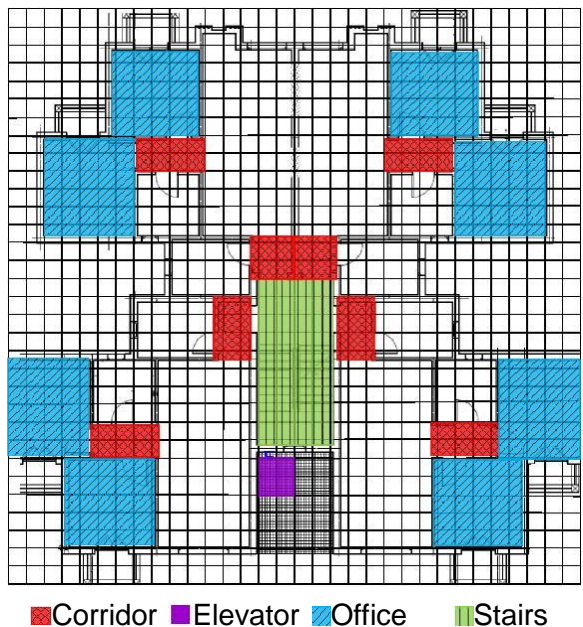

Figure 6: The floor plan used in our evaluation divided into blocks. Highlighted areas represent the different areas of interest. Note that the entire area is not highlighted for clarity.

In the rest of this section, we describe our features and the classifier used.

### 3.3.1 Classification Features

We use the following features to separate the different classes. These features are taken such that they are independent from the number of traces that go through the block. This is important for updating the floor plan as more traces become available.

- *Average time spent in block*: This feature represents the time spent per block averaged over all traces that go through the block. The intuition here is that users spend more time in different areas than other areas. For example, the average time spent in a block representing a corridor should be less than the average time spent in a block representing the elevator.

- *Average and variance of acceleration within the block*: These features represent the average and variance of the 3-axis accelerometer readings. For example, the average z-acceleration can be used to differentiate a stair from a corridor.
- *Correlation between accelerometer axes*: is especially useful for differentiating among activities that involve translation in just one dimension. For example walking through a corridor includes mainly one dimension. On the other hand, climbing stairs infer the additional Z dimension.
- *Average number of turns per trace*: This feature captures the number of turns inside a block per trace. It can be captured through the compass readings. This can help in identifying classes such as the elevator that has a specific turn pattern.
- *Average WiFi and GSM RSSI*: This can help in detecting classes where the RF signal is blocked, e.g. as in elevators.

### 3.3.2 Classifier

We used the C4.5 tree-based classifier [17] to build a decision tree from a training set to be used for classification. The C4.5 algorithm depends on the concept of information entropy based on divide-and-conquer strategy. A rule set can be produced in the form of if-then by traversing the decision tree.

We also use bootstrap aggregation ensemble learning [14] technique. This enables us to parallelize the technique, and hence become more suitable for the cloud. In addition, using the combined system for classification can increase the learning system generalization ability. Bootstrap aggregating, involves having each model in the ensemble vote with equal weight. In order to promote model variance, bootstrap aggregation trains each model in the ensemble using a randomly-drawn subset of the training set.

### 3.3.3 Implementation

Our classifier was implemented on a Dell Latitude E6510 with an Intel core i7 2.67 GHz processor and 8 GB RAM. The operating system used is Linux Fedora 13 distribution. The server runs the Apache Hadoop software framework that supports data-intensive distributed applications under a free license. It enables applications to work with thousands of nodes and petabytes of data. We used the Hadoop MapReduce framework along with Hadoop MapReduce plug-in for eclipse to develop our MapReduce based software. For building the C4.5 tree based classifier, we used the Weka API for java [2].

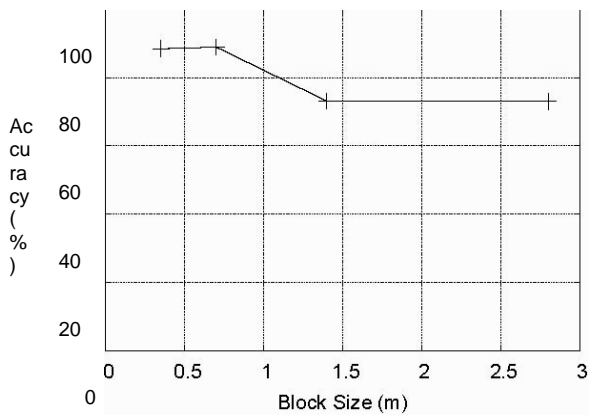

Figure 7: The effect of changing the block size on the classification accuracy.

| Actual Steps | FSM-based | Local Variance |
|---|---|---|
| 300 | 297 (1.0%) | 45 (85.0%) |
| 270 | 278 (2.9%) | 217 (19.6%) |
| 120 | 121 (0.8%) | 74 (38.3%) |
| 19 | 19 (0.0%) | 12 (36.8%) |
| 11 | 10 (9.0%) | 2 (81.8%) |
| 4 | 4 (0.0%) | 6 (50.0%) |

Table 2: The confusion matrix for the classifier for a block size of 0.7m.

### 3.3.4 Results

Figure 7 shows the effect of changing the block size on the accuracy of the system. The figure shows that a block size of 0.7m gives the best accuracy of 90%. Smaller block sizes makes less traces pass through the block while larger blocks cover more than one area. Both decrease the classification accuracy.

Table 2 shows the confusion matrix. The table shows that elevators and corridors can be detected with zero false negative rate. In addition, the elevators and offices have zero false positive rate. This highlights that elevators can be used as synchronization points as discussed in Section 3.2.1.

### 3.4 Discussion

In this section, we showed the feasibility of the floor plan generation module through experiments in a real environment. We presented algorithms for accurate generation of user traces based n identifying the number of steps. Our techniques can achieve a displacement error of less than 0.9m and less than 2.3% average error in estimating the number of steps. For area of interest identification, our results show that we can correctly identify the areas of interest with 90% accuracy. This accuracy can be further enhanced by applying higher level heuristics, for example by analyzing the adjacent tiles and removing noise.

For automating the fingerprint construction process, WiFi and GSM signatures can be stored in the fingerprint as the user traces are generated. The fingerprint information is updated as new traces are added. This captures the dynamic changes in the environment. Note that other localization technologies and sensed information, such as the accelerometer and compass data, can also be used to estimate the user location. Selecting the localization technology to use and balancing the accuracy/energy-consumption requirement is a direction for research in the IPS vision as we discuss in Section 4.6.

## 4 Challenges

In this section, we describe the different challenges that still need to be addressed to realize the ubiquitous indoor positioning system vision. In addition, we present different research directions related to the IPS problem.

### 4.1 Handling Hardware Variations

Since our vision is based on a crowd-sourcing approach, dealing with heterogeneous hardware is an unavoidable fact. This requires mapping between the signals collected by different sensors, using models that take into account the quality of the different sensors in different phones, designing dynamic algorithms to address the heterogeneity of power requirements, leveraging the different sensors on the phone to achieve mapping and localization, among others. Sensor fusion is a well developed area in signal processing while he other issues are still open for further research.

### 4.2 Seamless Roaming between Indoor and Outdoor Environments

A true ubiquitous IPS has to be seamlessly integrated with any outdoor positioning system to provide everywhere localization capabilities. This requires detection of roaming between indoor and outdoor environments, developing new applications to leverage everywhere localization (such as indoor/outdoor direction finding), developing new localization API's that are technology independent, along with addressing the scalability issues of a positioning system that covers the globe.

### 4.3 User Privacy

Whenever localization is discussed, user privacy becomes an issue. Although this problem is not unique to our system, it may generate more concerns as a ubiquitous IPS can provide a 24/7 user tracking. Defining user profiles that limit when location can be addressed, anonymizing the data and location, and asking for user permission are among the techniques that can be used to enhance the user privacy. Alternatively, downloading the fingerprint to the device and running the localization algorithm on the device is another approach to provide user privacy. New techniques that take into account scale of the system need to be investigated. Another issue related to privacy is that some sensitive buildings may not authorize the automatic construction of their floor plan. This can be addressed by requesting the user to authorize the data collection on the first time the system is activated in a certain building. Techniques for automating this task is a possible research direction. One approach can be to have specific information (e.g. in WiFi beacons) to automatically disable the system (similar to disabling web site robots). Techniques to reduce the accuracy of the system in certain buildings can also be investigated.

### 4.4 Robustness against Attacks

Since the system is open for data collection from all users, this creates a vulnerability for poisoning the reported data to reduce the system accuracy. The good news that the large number of users and data reduces the effect of attacks as it can be used to detect outliers.

### 4.5 Enhancing System Performance

Different techniques can be used to enhance the system performance. For example, exploiting the similarity between different floor plans in the same building can be used to reduce the data collected and to enhance accuracy. The number of floors can be estimated with high accuracy depending on the detected elevators and stairs. In addition, The amount of data collected from a specific building can be dynamically changed based on the number of traces already collected in this building and their coverage.

### 4.6 Energy-efficiency Aspects

Since our main data collection device is the cell phone, energy consumption becomes a major concern. The frequency of data collection and what sensors to turn on and off should be carefully studied. The IPS system need to be designed to carefully balance its accuracy and energy-efficiency requirements. Previous studies on the energy consumption of different phone components can be used in this regard [21]. However, such studies need to be expanded to a larger class of phones and to address more components.

### 4.7 Incentive Models

A crowd-sourcing-based system should have incentives to encourage the users to collect data for new buildings. This can be explicit, by given users credit based on the amount of data they contribute, or implicit by collecting data while the user is actually using the system and leveraging the times the system is left on. This is used, for example, in commercial systems such as Windows Live Search For Mobile and Google Maps for Mobile, where data is collected when the system is running (which can be disabled by the user).

### 4.8 Phone Orientation

The measured acceleration values are reported in terms of the local coordinates of the phone, not the world's coordinates. Therefore, the gravity will affect the measured acceleration in the 3-axes of the phone by an amount depending on the orientation of the phone. If such effect is not compensated for, it will severely affect the performance of the system. One way we handled this in our system is by depending on the magnitude of the acceleration for the step detection, rather than on the individual acceleration components. Another way is to use the gyroscope sensor, if available, to detect the phone orientation. New algorithms for step detection and/or areas of interest identification that are orientation-independent is a possible research direction.

### 4.9 Scalability Issues

As discussed in the previous subsections, scalability is a major concern in a true ubiquitous IPS that is expected to cover the entire globe. Techniques for data reduction to address the huge amount of data need to be developed. Clustering techniques to reduce the localization search space can also be investigated.

### 4.10 Other Technologies

In this challenge paper we showed the feasibility of the IPS system based on RF localization technologies (mainly WiFi and GSM). Although we believe that such technologies are the most promising, due to their ubiquitous installation, a true ubiquitous localization system should work with any available localization system, such as ultrasonic-based [16] or infrared-based localization [18]. The same techniques presented in this paper can be applied to these technologies. However, the main challenge lies in dealing with different localization technologies at the same time, both for indoor and outdoor environments and switching between them based on the user. The scalability issue is further exacerbated with the expansion of the supported technologies.

## 5 Related Work

Many systems over the years have tackled the indoor localization problem including the infrared [18], ultrasonic [16], and radio frequency (RF) [4,11,12,20]. All these systems are usually deployed in limited areas and assume the existence of building floor plans. All such technologies can be integrated as localization technologies in the global IPS vision as discussed in Section 4.10.

Recently, the idea of relying on user input, i.e. crowd-sourcing, to create a location database has been proposed for both outdoor and indoor localization, e.g. [5, 15]. Such systems focus on constructing the fingerprint database, but not on constructing the floor plan.

Simultaneous Localization and Mapping (SLAM) is a known technique in robotics for robot localization. However, contrary to what the name implies, it does not directly fit the ubiquitous IPS version. The most popular algorithms for solving the SLAM problem is based on the Extended Kalman Filter and the Rao-Blackwellized Particle Filters [8, 9]. In SLAM the location is estimated using the odometry data of the robot, where information about obstacles and landmarks are provided by using sensors, such as cameras and laser range detectors. Our work is different from SLAM in major points: (1) In SLAM the robot builds his own map of the area of interest from scratch. On the contrary, since our work is based on crowd-sourcing, the floor plan is estimated by processing multiple traces from different users. When ever a new user enters that area he/she can make use of the generated floor plan while at the same time contribute for its update. (2) The SLAM solution for mapping does not define areas of interest; it only specifies which areas are free to move through and which contain an obstacle or a landmark.

Similar in essence to our approach for indoor floor plan construction, a number of methods have been used to infer the street roadmaps from raw GPS traces, e.g. [6, 7] and collect GPS traces from users [1]. However, the challenges of both problems are significantly different. First, GPS traces accuracy is much more accurate than odometer readings for the required accuracy. Second, streets typically have directions, which can be leveraged to remove ambiguity and enhance accuracy. Third, estimated locations from the GPS samples are independent while the estimated locations using accelerometer samples are correlated in time. This makes the problem inherently noisier and requires new approaches. In addition, GPS is considered a ubiquitous outdoor positioning technology, while in this paper we target an equivalent indoor system.

## 6 Conclusions

In this paper, we introduced our vision for a ubiquitous indoor positioning system. An IPS system is envisioned to be able to provide indoor localization in a worldwide scale, with minimum overhead, to work with heterogenous devices, and to allow users to roam seamlessly from indoor to outdoor environments. Our approach is based on leveraging smart phones as ubiquitous computing and sensing devices along with the cloud for scaling the backend processing and storage capabilities. We described the IPS system's architecture and showed the feasibility of the automatic construction of a worldwide floor plan database. We presented algorithms for implementing different functionalities of the floor plan module of the IPS system that work with smart phones. In particular, we presented an algorithm for the construction of user traces from raw sensor data and a classifier to identify the different areas of interest in a building. Our results show that we can achieve a displacement error of less than 0.9m and less than 2.3% average error in estimating the number of steps. In addition, the system can achieve up to 90% accuracy for area of interest identification. Moreover, further processing can enhance this accuracy.

Our results have established the proof of feasibility of the IPS vision. In order to get it to the point that it can be commercially deployed we have to tackle a number of challenges. We identified different challenges for realizing the ubiquitous IPS system and give directions on how to address them.

We believe that pursuing our vision to ubiquitous localization systems will lead to major advances in the field: First, researchers and application developers will focus on the context-aware applications, rather than going into the details of the localization technology. Second, a new set of pervasive applications will be enabled as discussed in the paper. Third, such system will provide a rich environment to learn about systems to that scale to the degree to map the indoor buildings of the entire world, covering billions of devices and users. Fourth, new incentive models for users to participate in the distributed floor plan creation could be leveraged with other large scale pervasive systems. Fifth, the developed testbed created through the realizing of our vision can be used as a nucleus for a large scale heterogeneous pervasive testbed.